\documentclass[12pt]{article}
\usepackage[letterpaper, margin = 1in]{geometry}
\usepackage{amsmath, amsthm, amssymb, bbm}
\usepackage{graphicx, subcaption, caption, float}
\usepackage{booktabs, adjustbox}
\usepackage{hyperref, xcolor, bbm, setspace}
\usepackage{authblk}

\DeclareMathOperator*{\argmin}{arg\,min}
\DeclareMathOperator*{\argmax}{arg\,max}
\newcommand{\bm}[1]{\mbox{\boldmath$ #1 $\unboldmath}}
\usepackage{algorithm,algorithmic}
\usepackage{comment}
\usepackage{mathtools}

\usepackage[round]{natbib}
\title{\bf Adaptive Exploration and Optimization of Materials Crystal Structures}

\author[1]{Arvind Krishna}
\author[2]{Huan Tran}
\author[3]{Chaofan Huang}
\author[2]{\\Rampi Ramprasad}
\author[3]{V. Roshan Joseph\thanks{To whom the correspondence should be addressed: roshan@gatech.edu.}}

\affil[1]{Department of Statistics and Data Science, \protect\\ Northwestern University, Evanston, IL 60208}
\affil[2]{School of Materials Science and Engineering,  \protect\\ Georgia Institute of Technology, Atlanta, GA, 30332} 
\affil[3]{H. Milton Stewart School of Industrial and Systems Engineering, \protect\\ Georgia Institute of Technology, Atlanta, GA, 30332}

\begin{document}

\maketitle

\bigskip

\begin{abstract}

A central problem of materials science is to determine whether a hypothetical material is stable without being synthesized, which is mathematically equivalent to a global optimization problem on a highly non-linear and multi-modal potential energy surface (PES). This optimization problem poses multiple outstanding challenges, including the exceedingly high dimensionality of the PES and that PES must be constructed from a reliable, sophisticated, parameters-free, and thus, very expensive computational method, for which density functional theory (DFT) is an example. DFT is a quantum mechanics based method that can predict, among other things, the total potential energy of a given configuration of atoms. DFT, while accurate, is computationally expensive. In this work, we propose a novel expansion-exploration-exploitation framework to find the global minimum of the PES. Starting from a few atomic configurations, this “known” space is expanded to construct a big candidate set. The expansion begins in a non-adaptive manner, where new configurations are added without considering their potential energy. A novel feature of this step is that it tends to generate a space-filling design without the knowledge of the boundaries of the domain space. If needed, the non-adaptive expansion of the space of configurations is followed by adaptive expansion, where “promising regions” of the domain space (those with low energy configurations) are further expanded. Once a candidate set of configurations is obtained, it is simultaneously explored and exploited using Bayesian optimization to find the global minimum. The methodology is demonstrated using a problem of finding the most stable crystal structure of Aluminum.

\end{abstract}

\noindent%

\bigskip

\section{Introduction}

One of the most ambitious goals of material scientists is to discover and design new materials with desirable properties and applications \citep{franceschetti1999inverse, weymuth2014inverse, d2012genetic, xiang2013towards, huan2015accelerated, Arun:design}. Until the present time, material discoveries are largely driven by expensive and time-consuming trial-and-error approaches, i.e., they must be physically synthesized and tested in a laboratory with limited guidance beyond empirical rules and experience. However, under some scenarios, some properties of a material can be computed without synthesizing it, if its  atomic structure is known. 

Predicting the stable atomic configurations of a given set of atoms can be mathematically formulated as an optimization problem. The most stable configuration is the global minimum of the potential energy surface corresponding to all possible atomic configurations. This is a very active research area in the emerging era of materials discovery and design, when a large number of hypothetical materials should be examined by computational methods before some of them can be advanced to the synthesizing and testing steps. The main objective of materials structure prediction \citep{oganov2019structure} is searching for low-energy atomic configurations of a given set of atoms. 

Although we are primarily interested in the global minimum of the PES, certain local minima may also be useful \citep{HuanData2021,therrien2021metastable}. External perturbations such as temperature, pressure, and other kinetic-related factors may bring a local minimum down to be the global minimum at a specific condition \citep{huan2018pressure, kobayashi2019formation,  gaida2021phase}, or drive the atomic configuration to land at some nearby (accessible) local minima. Therefore,  configurations that are very far from (and/or very well-separated by a high potential energy barrier with) the global minimum may also be reliable. 

The specific class of materials addressed in this work is crystal. A crystal can be imagined as an infinitely repeated array of a unit cell along three Cartesian dimensions.  Crystal materials are dominant in material science because of two main reasons. First, a majority of materials are  crystals and/or can be modeled very well by crystal models. Second, because of its periodicity, crystal models are small enough so that physics-based computational methods such as the Density Functional Theory (DFT) \citep{DFT1,DFT2}, the most reliable (but expensive) parameter-free computational method, may be used at an acceptable cost. 

The least biased and non-empirical approaches to crystal structure prediction involve computational optimization \citep{oganov2011modern}. These approaches involve explicit computation of the potential energy of the crystal structure, followed by solving an optimization problem to find the crystal structure corresponding to the least energy, or the thermodynamically most stable configuration. \cite{pickard2006high,pickard2011ab} developed a random-search based method to find the stable crystal structure configuration. The underlying idea in this method is to use the DFT to optimize a randomly generated set of crystal structures, driving each of them to the nearest local minimum. With a large number of randomly generated samples, these approaches can successfully identify the most stable crystal structure configuration in many cases. Some other recently developed methods are simulated annealing \citep{pannetier1990prediction,schon1996first, tekin2010first}, basin hopping \citep{wales1997global}, minima hopping \citep{goedecker2004minima}, metadynamics \citep{martovnak2006crystal}, evolutionary algorithms \citep{trimarchi2009predicting}, and USPEX (Universal Structure Predictor: Evolutionary Xtallography) approach \citep{oganov2006crystal,glass2006uspex,oganov2011evolutionary}, which is based on evolutionary algorithms. 
 
While these methods are different in many aspects, most notably the employed (global) optimization algorithms, they do share two common fundamental problems. First, given a set of atoms, how to thoroughly explore the configurational space, and second, within this accessible domain, how to efficiently identify the global minimum? Since the number of local minima of the potential energy scales up exponentially with the number of atoms in the system \citep{berry1993potential,stillinger1999exponential}, both problems are enormously challenging. In most of the practical cases, there is essentially no way to guarantee that the entire configurational space can be explored, and for this reason, new developments in this active research area are still in progress.
 
We have developed an expansion-exploration-exploitation framework to address these problems, i.e., (i) enlarging the accessible domain of the search space, and (ii) finding the global minimum of the PES within this domain. For the first problem, we expand the space spanned by a few possible configurations by perturbing them and generating more configurations in their neighborhood. The configurations are generated such that they continuously expand the spanned domain space of configurations, especially towards the low-energy regions of the domain space. Once a representative candidate set of configurations is obtained, a Bayesian optimization procedure \citep{jones1998efficient} is used for exploring the domain space regions with high uncertainty in the potential energy estimate while simultaneously exploiting the low-energy regions to find the global minimum and reliable local minima among the candidate set of configurations.

\begin{figure}[h]
\begin{center}
\includegraphics[scale=0.35]{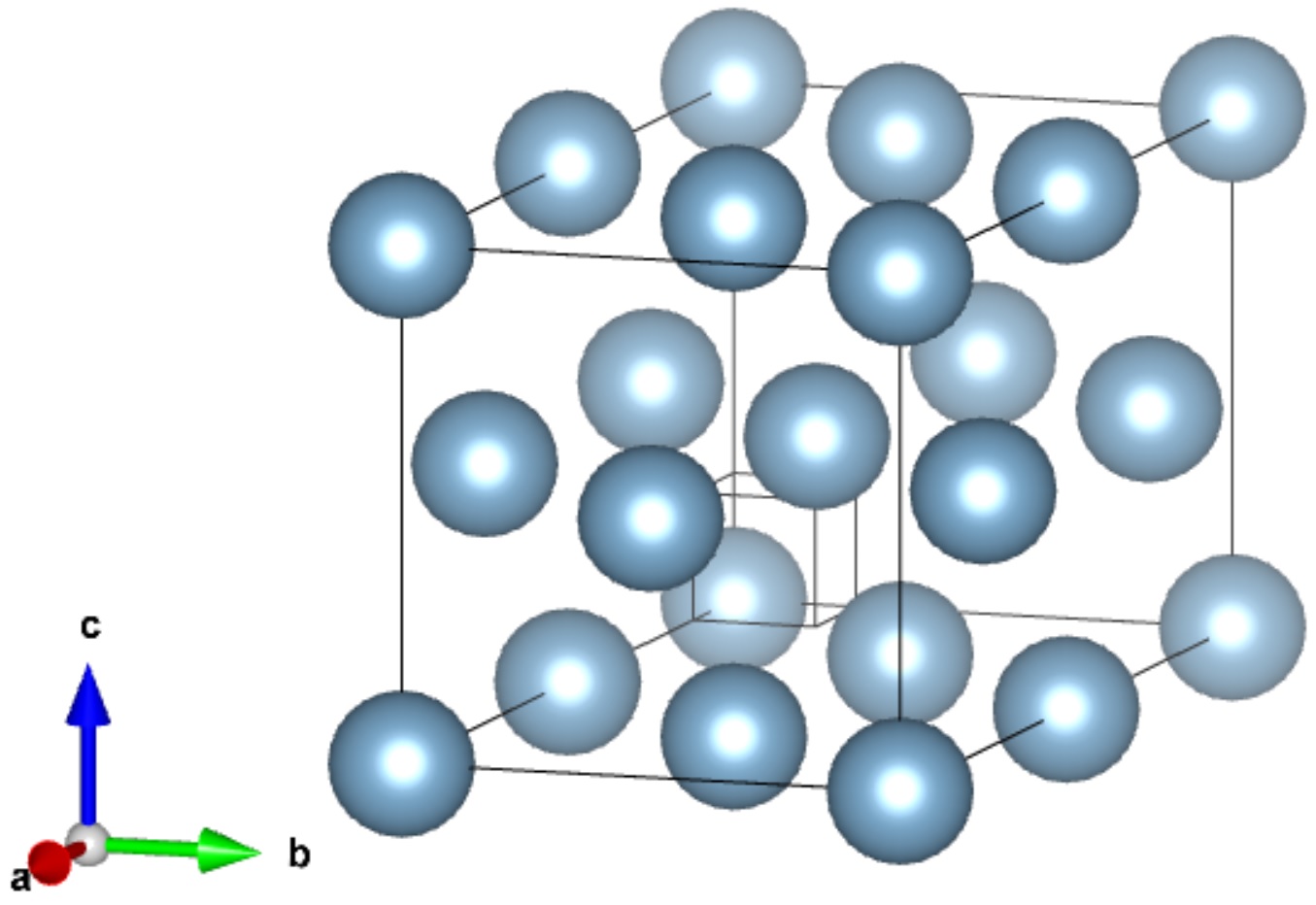}
\caption{A supercell of the body-centered cubic (ground state) crystal structure of Aluminium. The ideal crystal is obtained by infinitely repeating this cell in three dimensions.}
\label{fig:AL_true_structure}
\end{center}
\end{figure}

This article is organized as follows. In Section 2, we formulate the crystal structure, discuss its representation and energy computation using the DFT. In Section 3, we mention the constraints and challenges of the problem. In Section 4, we describe the developed methodology that addresses these constraints and challenges. In Section 5, we illustrate the effectiveness of our methodology on the problem of finding the crystal structure of $Al_8$ (Aluminum), where the true structure is already known (see Figure \ref{fig:AL_true_structure}). We conclude the article with some remarks in Section 6.

\section{Materials crystal model}
This section describes the crystal structure and the computation of its potential energy. 

\subsection{Parameters of a crystal structure}
A crystal model of a material includes a parallelepiped unit cell defined by three basis vectors $\vec a$, $\vec b$, and $\vec c$, a given set of $N_A$ atoms arranged in the unit cell, and an assumption that the unit cell is infinitely repeated along $\vec a$, $\vec b$, and $\vec c$.  Figure \ref{fig:AL_true_structure} shows an example of a unit cell. Because a material does not change under rigid translations and rotations, three vectors $\vec a$, $\vec b$, and $\vec c$ can be uniquely determined by six independent numbers. Therefore, the crystal structure prediction is mathematically equivalent to a global optimization problem on the PES defined in a $3N_A + 6$ dimensional space ($N_A$ has no upper limit, and its typical values can be as high as $100$).

\subsection{Computing the potential energy of a crystal structure: DFT}
We use single-point DFT computations to determine the potential energy of the atomic configuration under consideration. Such calculations are performed using the {\sc abinit} package \citep{Gonze_Abinit_3}, employing the Perdew-Burke-Ernzerhof functional \citep{PBE} for the quantum mechanical exchange-correlation energies. The electron-nuclear interactions are computed with help from the norm-conserving Hartwigsen-Goedecker-Hutter pseudopotentials \citep{HGH_Pseudo}. For our calculations, the Brilouin zone is sampled by a dense Monkhorst-Pack k-point mesh \citep{monkhorst}, and a basis set of plane waves with kinetic energy up to 550 eV.

\section{The Problem: Constraints and Challenges}

\begin{figure}[h]
\begin{center}
\includegraphics[scale=0.45]{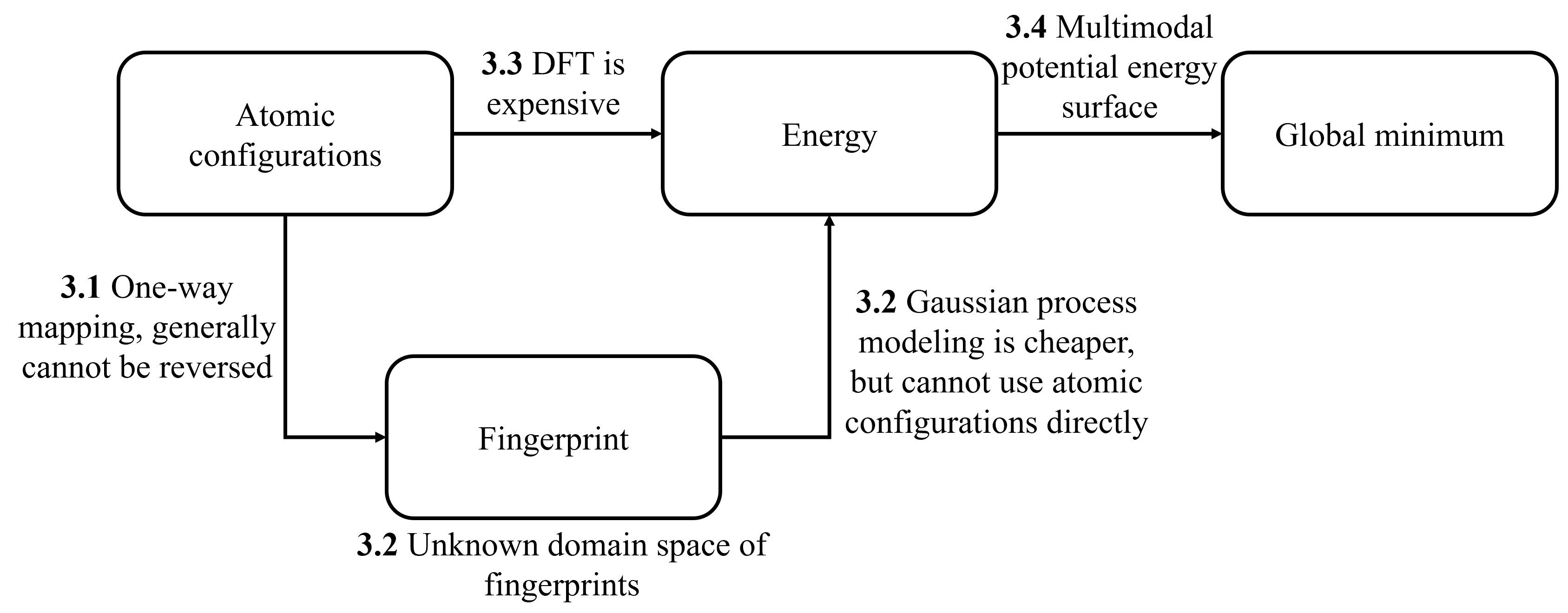}
\caption{Constraints and challenges in the problem.}
\label{fig:constraints}
\end{center}
\end{figure}

The constraints and challenges of the problem are depicted in Figure \ref{fig:constraints} and are explained in detail in this section.

\subsection{Crystal structure representation: One-sided mapping}
As mentioned earlier, we use the DFT to compute the potential energy of a crystal structure. Potential energy of a crystal structure depends only on the relative distance between its atoms, and not on the absolute positions of atoms. This implies that the energy is invariant to translational, rotational and permutational operations of alike atoms, in the crystal structure configuration. Such transformations change the Cartesian coordinates of the atom, but do not change the material in any physical and chemical way. For this reason, we used the Cartesian coordinate system for energy computations with DFT, while using the \textit{AGNI} (Adaptive, Generalizable and Neighborhood Informed) fingerprint \citep{batra2019general} to map the atomic configurations onto their energy. All the redundant Cartesian coordinate system atomic configurations correspond to a unique \textit{AGNI} fingerprint. An \textit{AGNI} fingerprint captures the atomic-level information of the structure pretty well while preserving the material presentation under such ``identity'' transformations in the materials space. 

The \textit{AGNI} fingerprint used in this work is defined as $f \coloneqq \{S_k;  V_k\}_{k=1}^n$, where the scalar components $S_k$ and the vectorial components $V_k$ are given by
\begin{equation}
S_k = \sum_{i\not= j} {\cal G}(r_{ij},\sigma_k)f_c(r_{ij}),
\end{equation}
and 
\begin{equation}
    V_k = \sqrt{\sum_{\alpha=x,y,z}\left[\sum_{i\not= j} \frac{r_{ij}^\alpha}{r_{ij}}{\cal G}(r_{ij},\sigma_k)f_c(r_{ij})\right]^2},
\end{equation}
respectively. Here, $r_{ij}$ is the distance between atoms $i$ and $j$, $r_{ij}^\alpha$ is the projection of $r_{ij}$ onto the Cartesian axis $\alpha$, ${\cal G}(r,\sigma_k)$ is the Gaussian function centered at 0 with varying width $\sigma_k$:
\begin{equation}
    {\cal G}(r,\sigma_k) = \frac{1}{\sqrt{2\pi}\sigma_k}\exp\left(\frac{-r^2}{2\sigma_k^2}\right),
\end{equation} and $f_c(.)$ is a cutoff function that is used for disregarding interaction among atoms that are further than a distance $R_c$ from each other:
\begin{equation}
    f_c(r)\equiv \frac{1}{2}\left[\cos(\pi r/R_c)+1\right].
    \label{eq:fc}
\end{equation}
While summarizing over the atoms $i$ and $j$, the periodicity of the unit cell is considered, i.e., for an atom, its neighbors in all the repeated images of the unit cell are also taken into account. The cutoff function, $f_c(r)$, defined in \eqref{eq:fc} is used for restricting the neighborhood to a radius of $R_c$. We used $R_c = 8$ \AA ~ in this work, because the interaction between two atoms at this distance is negligible. From the mathematical point of view, \textit{AGNI} fingerprint is a way of projecting the atomic positions onto a set of predefined basis functions. Here, we used $n=16$ functions ${\cal G}(r,\sigma_k)$, thus our fingerprint $f$ has $2n=32$ components or dimensions. Note that the accuracy of the model using a fingerprint increases as the dimension increases and then saturates. Our tests indicate that after 32 dimensions, the increase in model accuracy with increasing dimensions is negligible. The \textit{AGNI} fingerprint is one of the numerous material fingerprints \citep{behler2007generalized, bartok2010gaussian} developed during the last decade. 

Although the $AGNI$ fingerprint eliminates redundancy in the Cartesian coordinate representation of the crystal structure configuration, and reduces the PES dimensionality from $3N_A+6$ (as described in Section 2) to $32$, it introduces a constraint. We can only map a configuration in the Cartesian coordinate system to the $AGNI$ system, but not vice-versa.

In the absence of the above constraint, we could have used a continuous optimization procedure in the \textit{AGNI} fingerprint space to find a solution and transform it to the physically interpretable Cartesian coordinate system. However, in the presence of this constraint, we will need to use a discrete optimization approach of considering a candidate set of configurations, finding their corresponding fingerprints, and then finding the one with the least energy. This gives rise to the challenge that the candidate set of fingerprints must contain the solution(s) or fingerprints ``close enough'' to the solution(s).

\cite{tripathy2016gaussian, siivola2021good, wang2022scalable} develop the Gaussian process model and optimize it in the latent space. However, we cannot use these methods as there is no reverse mapping from the \textit{AGNI} fingerprint space (latent space) back to the original input space (Cartesian coordinate system). \cite{chen2020semi} proposed formulating a least square problem to map the solution back to the input space. However, this becomes a discrete optimization problem in our case, which is computationally very expensive to solve.

\subsection{Crystal structure representation: Unknown domain space}
The domain space of the crystal structure configuration is more intuitive than that of the \textit{AGNI} fingerprint. This gives rise to the challenge of obtaining a candidate set of fingerprints that are representative of all fingerprints, or a candidate set of crystal structure configurations that are representative of all possible configurations. If we knew the domain space, we could have used a space-filling design \citep{joseph2016space} to obtain a representative candidate set of fingerprints. However, the challenge is to find a representative set of fingerprints without the knowledge of their domain space.

There has been some work done for performing Bayesian optimization in an unknown input space \citep{shahriari2016unbounded,nguyen2017bayesian,ha2019bayesian}. However, these methods are not applicable to our problem due to several reasons. First, these methods systematically expand the input space to search for the optimum. But, in our problem, we do not expand the input space (i.e., the Cartesian coordinate system space) directly. We map the input space to a feature space (the \textit{AGNI} fingerprint space), which is expanded systematically to search for the optimum. Second, we cannot avoid the input space, and apply these methods directly on the feature space because a solution in the feature space cannot be mapped back to the input space, and the solution is to be found in the input space. Third, we cannot avoid the feature space and apply these methods on the input space. With the large amount of redundancy in the input space, the number of minima will be too large making the problem unnecessarily complex.
There has been some work done to identify the feasible domain space when the search space is unknown. \cite{basudhar2008adaptive} used Support Vector Machines, while \cite{chen2017beyond} used active learning to identify the feasible domain in an unbounded space. However, in our problem, it is not useful to identify the feasible feature space because the points in this space cannot be mapped back to the input space. Even though we expand the feature space, the expansion must be driven from the input space so that we can trace the solution in the feature space back to the input space.

\subsection{Expensive energy computation: Density functional theory (DFT)}
To find the most stable crystal structure configuration, we need to find the one with the least potential energy. As mentioned earlier, any reliable predictions of materials structure must be done with accurate-enough methods to compute the potential energy, and DFT is possibly the least expensive method of this kind. Even with this method, each evaluation of the potential energy on supercomputers requires hours or even days, depending on the size of the atomic system under investigations. This is why predicting a simple crystal structure by computations was regarded as ``{\it one of the continuing scandals in the physical sciences}'' in 1988 by a Nature's editor, Sir John Maddox \citep{maddox1988crystals}. Although structure prediction methods have evolved dramatically since then and have led to numerous new materials predicted computationally and realized experimentally \citep{oganov2019structure}, this remains a major bottleneck of contemporary materials discoveries. The expensive DFT computations constrain us to evaluate the energy for only a few configurations, which gives rise to the challenge of optimizing a huge potential energy surface, while observing it at only a few points. 

\subsection{Multi-modal potential energy surface (PES)}
The potential energy surface is highly nonlinear and multi-modal. Given that we can observe it at only a few points (see constraint $3.3$), it becomes challenging to accurately model all the modalities. As the number of local minima scale up exponentially with the number of atoms in a unit cell, $N_A$, the challenge is even bigger for crystal structures with large $N_A$. However, modeling the multi-modalities is necessary to find the global minimum as well as other reliable local minima.

\section{Methodology}
We have developed an expansion-exploration-exploitation framework for crystal structure prediction that addresses all the challenges presented in Section 3. This framework is implemented in two steps. The first step is \emph{domain space expansion}, where we expand the space spanned by a few possible configurations by iteratively adding more configurations. This leads to a candidate set of configurations that will ideally either span the entire domain space of possible configurations or at least span the space of stable configurations. The expansion step consists of a sequence of two sub-steps: \emph{non-adaptive expansion} and \emph{adaptive expansion}. Non-adaptive expansion refers to adding configurations without considering their potential energy. This tends to include unexpected configurations in our candidate set. If needed, this step is followed by adaptive expansion, which tends to add configurations that further expand the low-energy regions of the domain space. The expansion step is followed by simultaneous exploration and exploitation of the domain space spanned by the candidate set to find the configuration that corresponds to the minimum potential energy. We will explain these steps in the three sub-sections below.

In the sub-sections below, the number of initial configurations are denoted as $n_0$, the number of configurations added in the non-adaptive and adaptive expansion steps are denoted as $n_1$ and $n_2$ respectively. The cumulative number of configurations in the candidate set at the end of the non-adaptive and adaptive expansion steps are denoted as $N_1$ and $N_2$ respectively. The number of iterations in the Bayesian optimization procedure is denoted as $n_3$.

\subsection{Non-adaptive domain space expansion}
The purpose of this step is to obtain a candidate set of configurations that span as much domain space as possible. We start from a set of few possible configurations, and iteratively add those configurations to the set that expand their spanned domain space. The potential energy of the configurations is ignored, while developing the candidate set, to serve two purposes. First, it may lead us to regions of the domain space where a low-energy configuration is unexpected. Second, it saves the computational resources for calculating the energy and helps us obtain a larger candidate set within a given time period.

We will explain the algorithm with a toy example.  Let the  fingerprint domain space, be $[-3,3] \times [-3,3]$. However, in practice, we are not aware of the fingerprint domain space. So, we will not feed this domain space to our algorithm. Nevertheless, the objective of our algorithm will be to find a candidate set of fingerprints that fill this  space.

\begin{figure}[h]
\begin{center}
\includegraphics[scale=0.35]{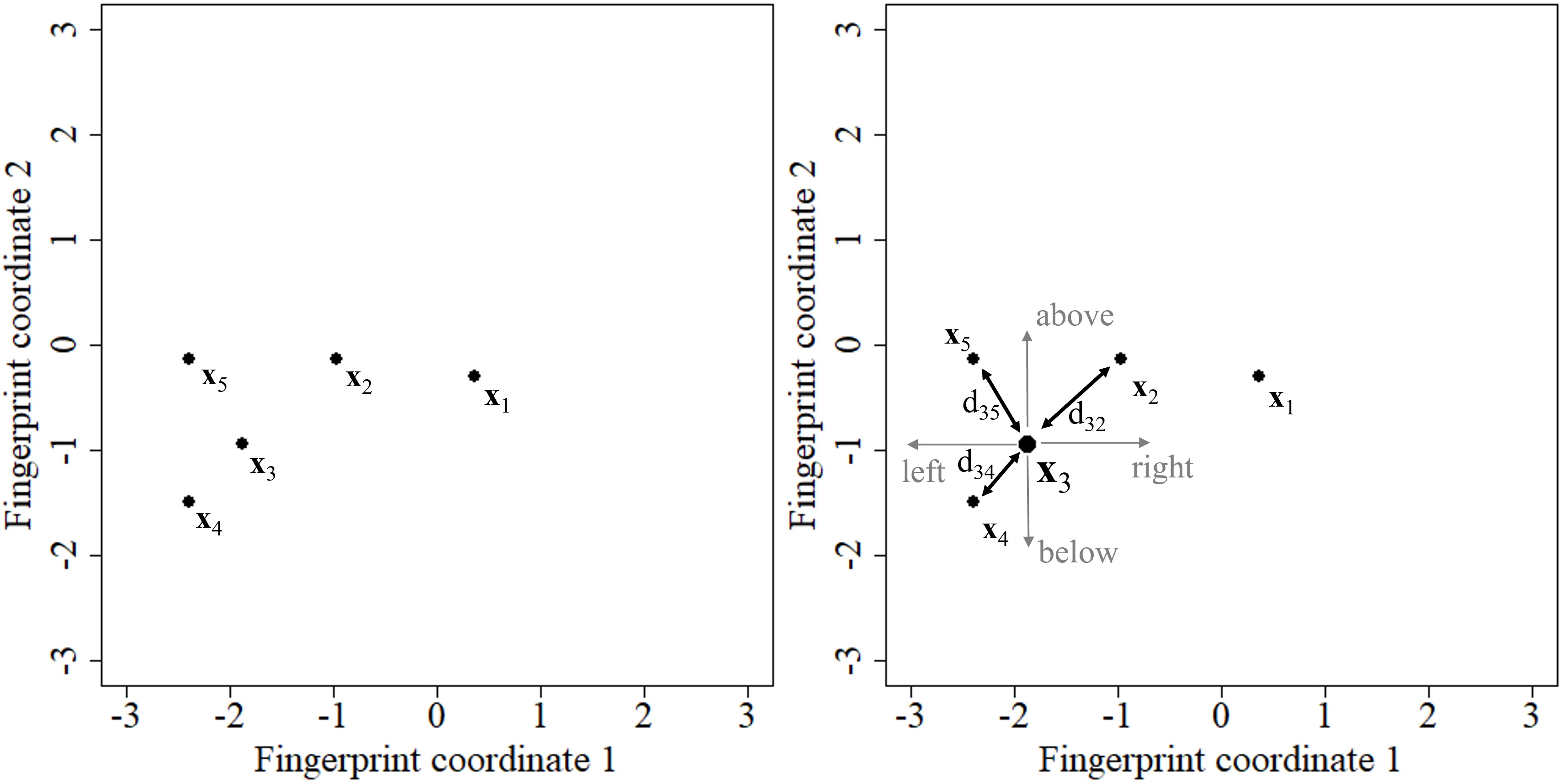}
\caption{(left): Initial $n_0=5$ fingerprints; (right): Distances to the closest neighbors of fingerprint $\bm x_3$ in all of its $2p=4$ neighborhoods.}
\label{fig:ini_fp}
\end{center}
\end{figure}

The algorithm begins by considering the initial candidate set of few possible atomic configurations, say $\mathcal C = \{\bm c_1,\cdots,\bm c_{n_0}\}$, where $\bm c_1,\cdots,\bm c_{n_0}$ are the $n_0$ initial atomic configurations. Let their corresponding fingerprints be $\mathcal X=\{\bm x_1,\cdots,\bm x_{n_0}\}$. Let us assume that there are a set of $n_0=5$ possible fingerprints for our toy example, as shown in Figure \ref{fig:ini_fp} (left), and we have a budget of expanding it to $N_1$ fingerprints. 

To expand the domain space spanned by the fingerprints, we will identify the most sparsely populated region of the domain space, and generate fingerprints around it. We define the most sparsely located fingerprint as the one that has the farthest nearest neighbor in any of its neighborhoods. A fingerprint-neighborhood is defined as the space on either side of the fingerprint along each  dimension. Thus, for a $p$-dimensional fingerprint, there are $2p$ neighborhoods - two on either side of it along each dimension. We intend to identify the fingerprint that has the farthest nearest neighbor in any of its $2p$ neighborhoods.

Consider the two-dimensional fingerprints in Figure \ref{fig:ini_fp}. Let us find the distance to the farthest nearest neighbor of fingerprint $\bm x_3$ in Figure \ref{fig:ini_fp} (right). For that we will find the distance to the nearest neighbor in $2p=2 \times 2 = 4$ neighborhoods - above and below $\bm x_3$, and right and left of $\bm x_3$. The nearest neighbors above and below are $\bm x_3$ are at a distance of $d_{35}=0.96$ and $d_{34}=0.75$ respectively, while those on the left and right are at distance of $d_{34}=0.75$ and $d_{32}=1.22$ respectively. Thus, the farthest nearest neighbor of $\bm x_3$ is at a distance of $\max(d_{32},d_{34},d_{35}) = 1.22$. 

\begin{figure}[h]
\begin{center}
\includegraphics[scale=0.35]{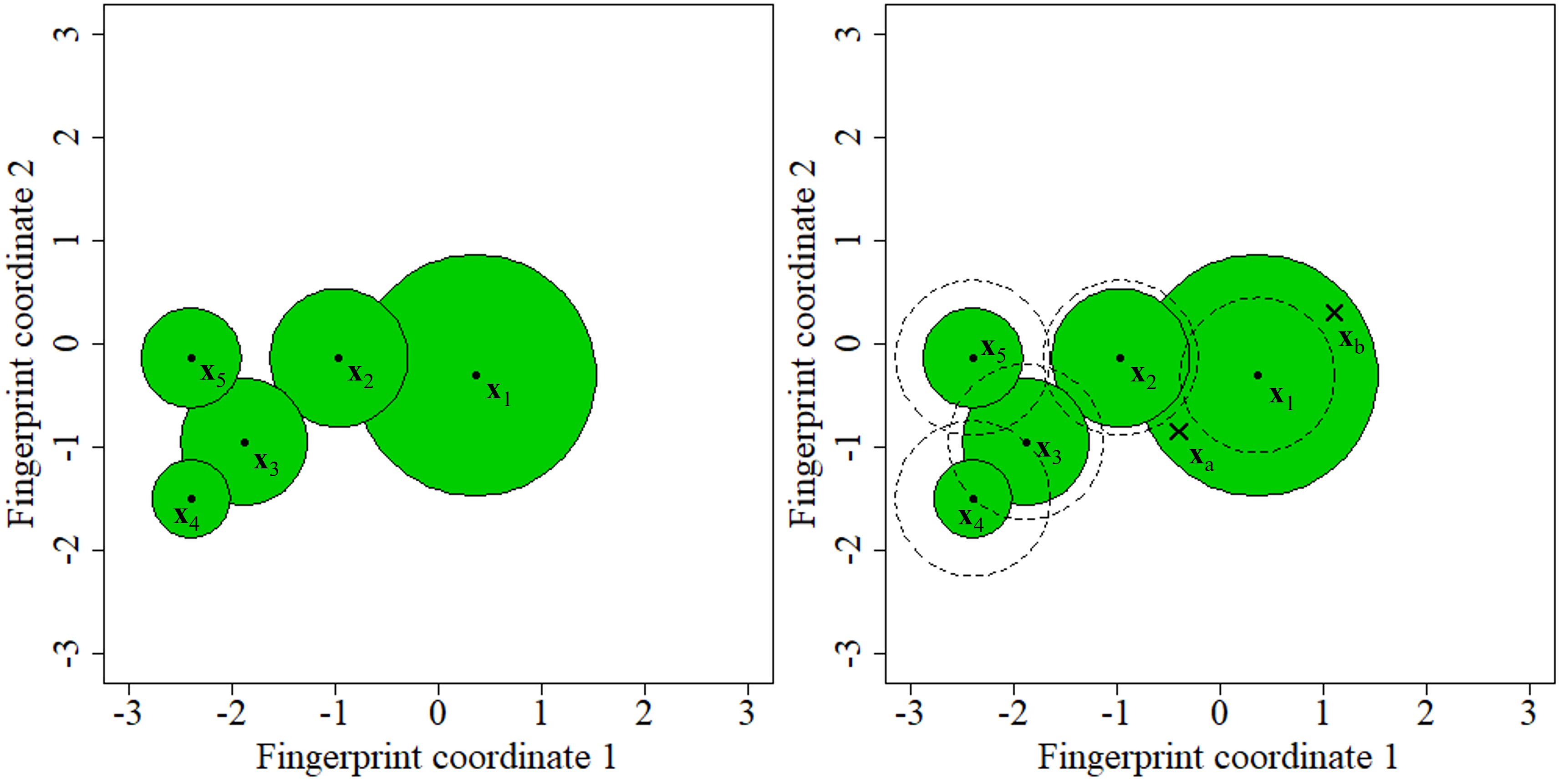}
\caption{(left): Space spanned by the initial $n_0=5$ fingerprints; (right): Dotted circle around each fingerprint with radius $t$, showing the minimum distance necessary between them and an acceptable newly generated fingerprint; Two examples of acceptable new fingerprints for inclusion in the candidate set - $\bm x_a, \bm x_b$.}
\label{fig:maxd_fp}
\end{center}
\end{figure}

Figure \ref{fig:maxd_fp} (left) visualizes the distance to the farthest nearest neighbor with a circle having radius half of that distance, around each fingerprint. Clearly, the fingerprint $\bm x_1$ is the most sparsely located fingerprint.  Let us label the fingerprint $\bm x_1$ as $\bm x_{sparse}$. So, we will find a fingerprint in the space around $\bm x_{sparse}$, and add it to the candidate set to expand the spanned domain space. 

To generate a fingerprint around $\bm x_{sparse}$, we randomly perturb the atomic configuration corresponding to $\bm x_{sparse}$ to generate another configuration, and fingerprint it. This new fingerprint is added to the candidate set if it is ``far enough'' from its nearest neighboring fingerprint. We use a {threshold distance} $t_i$  (for the $i{th}$ iteration), which will be defined later, to check if the randomly generated fingerprint is ``far enough''. If it is not ``far enough'', then we discard it, and again perturb the same atomic configuration. 

The purpose of the {threshold distance} $t_i$ is twofold. First, it is used for avoiding redundancy of fingerprints in the candidate set. Second, it ensures that the fingerprints are evenly spaced-out in their domain space. If $t_i$ is the threshold distance in the $ith$ iteration, and $d_{min,i}$ is the distance of the new fingerprint to its nearest neighbor in this iteration, then the threshold distance for the next iteration is given by:
\begin{equation}\label{eq:threshold_dist}
t_{i+1} = \frac{t_i+d_{min,i}}{2}, \ \forall i>1.
\end{equation}
The term $d_{min,i}$ ensures that the threshold distance is large when large parts of the domain space are unexplored, and small if the domain space is already well-explored. This makes the fingerprints spread farther apart until the entire domain space has been explored. Once the domain space has been explored, the threshold distance decreases so that new fingerprints may be added to the candidate set, until the budget of $N_1$ fingerprints is exhausted. The term $t_i$ ensures that the threshold distance does not change abruptly for an abrupt change in $d_{min,i}$. For the first iteration, $i=1$, $t_1$ is taken as the mean of the distances to the nearest neighboring fingerprint for each fingerprint.

In our toy example, $\bm x_1$ is the fingerprint identified with the farthest nearest neighbor. So, we perturb the atomic configuration corresponding to it to generate another one, and fingerprint it.
In all the examples of this Section, we perturb a fingerprint using a multivariate normal distribution. The mean of the distribution is the coordinates of the fingerprint being perturbed, and the covariance is a diagonal matrix, where the diagonal elements are the mean distance of a fingerprint to its nearest neighbor in the initial candidate set. Figure \ref{fig:maxd_fp} (right) shows the five fingerprints in the candidate set with a dotted circle around them whose radius is equal to the threshold distance $t_1=0.75$. If the new fingerprint falls inside any of the dotted circles, then it will be rejected on account of being redundant with the fingerprints in the candidate set. Figure \ref{fig:maxd_fp} (right) shows two examples of an acceptable new fingerprint - $\bm x_a$ and $\bm x_b$. Both of them expand the space spanned by the candidate set of fingerprints.

\begin{figure}[h]
\begin{center}
\includegraphics[scale=0.3]{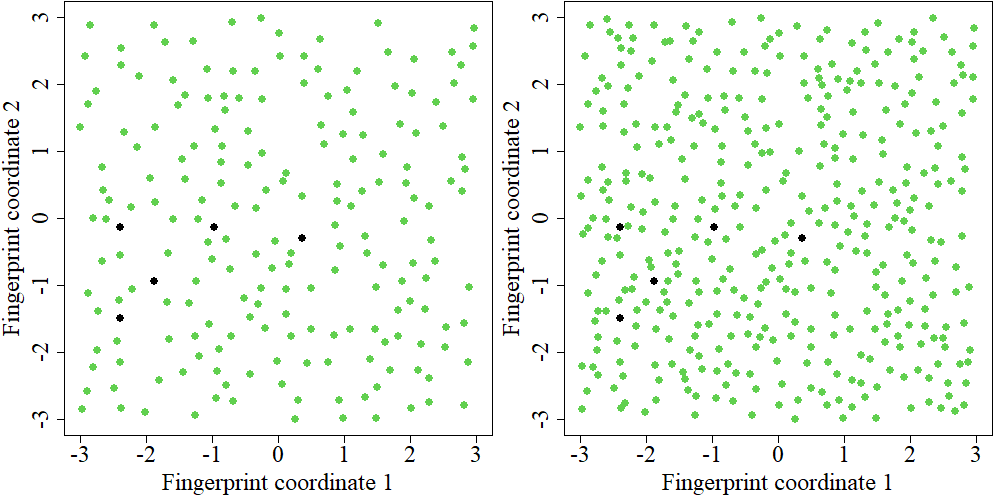}
\caption{Candidate set of $N_1=200$ fingerprints (left) and $N_1=400$ fingerprints (right) in the toy example.}
\label{fig:final_fp}
\end{center}
\end{figure}

We repeat the exercise of identifying and adding a fingerprint around the most sparsely located one, until we have added the desired number of fingerprints in the candidate set. Figure \ref{fig:final_fp} shows the results obtained when we have a candidate set of $N_1=200$ fingerprints, and $N_2=400$ fingerprints. Our algorithm performs well in (a) providing a candidate set of fingerprints that spans the entire domain space, (b) spacing-out fingerprints such that they evenly span the domain space within a given budget of $N_1$ fingerprints. The non-adaptive domain space expansion algorithm is included as Algorithm 1. 

\begin{algorithm}
  \caption{: Non-adaptive domain space expansion}
  \begin{algorithmic}[1]
  \STATE Input $N_1, \mathcal{C}, \mathcal{X}$
  \STATE $i \leftarrow n_0$
  \STATE Compute $\bm R= \{r_1, \cdots, r_{n_0}$ \COMMENT{Distance to farthest nearest neighbor among $2p$ neighborhoods, for each fingerprint }
  \STATE Compute $\bm D= \{d_1, \cdots, d_{n_0}$ 
  \COMMENT{Distance to the nearest neighbor of each fingerprint}
  \STATE $t \leftarrow$ mean(\bm D)
  \WHILE{$i \le N_1$}
  \STATE per $\leftarrow \argmax_{i}{r_i}$
  \STATE Randomly perturb $\bm c_{per}$ to generate $\bm c_{new}$
  \STATE $\bm x_{new} \leftarrow$ fingerprint$(\bm c_{new})$
  \STATE Compute $d_{min}$ \COMMENT{Distance to the nearest neighbor of $\bm x_{new}$}
  \IF{$d_{min}>t$}
  \STATE $\mathcal C \leftarrow$ append($\mathcal C,\bm c_{new}$)
  \STATE $\mathcal X \leftarrow$ append($\mathcal X,\bm x_{new}$)
  \STATE $i \leftarrow i+1$
  \STATE Update $\bm R, \bm D$
  \ENDIF
  \STATE $t \leftarrow 0.5(t+d_{min})$
  \ENDWHILE
  \STATE Output $\mathcal C, \mathcal X, \bm R, \bm D, t$
  \end{algorithmic}
\end{algorithm}

Regarding the choice of $N_1$, ideally,  it should be the number of fingerprints that are sufficient to capture the global minimum. However, since it is impossible to know this beforehand, we suggest a candidate set of size $N_1 = 100p$, as a thumb-rule, for a $p$-dimensional fingerprint. A larger $N_1$ could be used, but it will increase the computational and storage costs as discussed at the end of Section 4.3.

\subsection{Adaptive domain space expansion}
Adaptive domain space expansion is an optional step. It is not needed if the candidate set, obtained with the non-adaptive expansion algorithm, captures the global minimum. Although it is impossible to determine if the global minimum has been captured, it is possible to make an educated guess if further expansion of a particular region of the explored domain space may help capture more minima, one of which may potentially be the global minimum.

\begin{figure}[h]
\begin{center}
\includegraphics[scale=0.35]{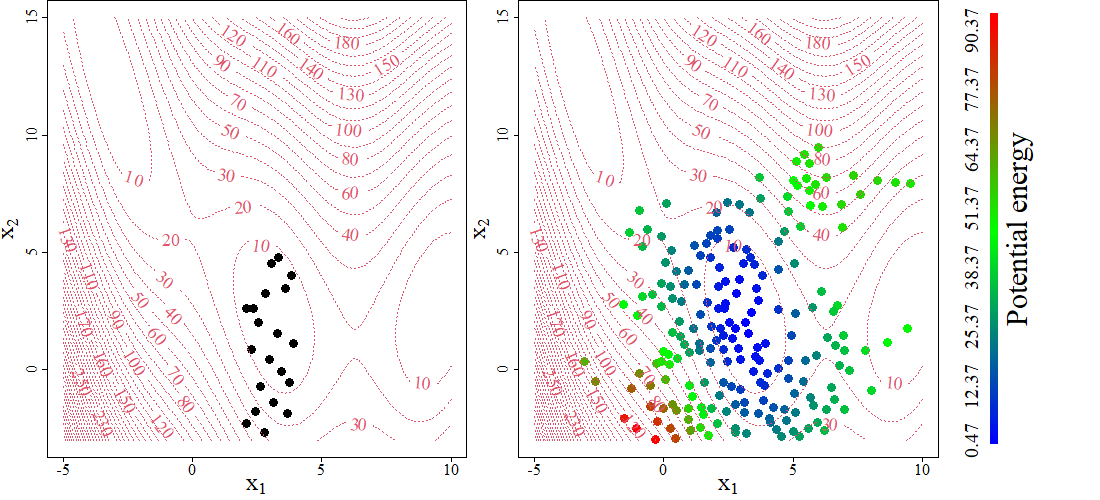}
\caption{Candidate set of  $n_0=20$ initial fingerprints (left) and  $N_1=200$ fingerprints obtained from the non-adaptive expansion algorithm (right), shown over the contour plot of the Branin function.}
\label{fig:branin1}
\end{center}
\end{figure}

To illustrate the need for adaptive expansion, we will consider an example, where it is assumed that the fingerprints are two-dimensional and their potential energy is given by the Branin function \citep{simulationlib}. Assume that there are $n_0=10p=20$ initial fingerprints as shown in Figure \ref{fig:branin1} (left). We use the non-adaptive expansion algorithm to expand them to a set of $N_1=100p=200$ fingerprints as shown in Figure \ref{fig:branin1} (right). As the budget uptil the non-adaptive expansion ($N_1=200$) step is exhausted, we will shift the focus to only the ``promising regions'' or the low-energy regions of the domain space, instead of the sparsely populated regions. We use DFT to compute the energy of the initial $10p=20$ fingerprints. Then, a model-based approach will be used for identifying the low-energy ``promising regions''. 

As the $n_0$ fingerprints corresponding to the initially known configurations are likely to be stable, we compute their potential energy. If $n_0<10p$, we suggest to choose the remaining fingerprints from the set of $N_1$ fingerprints, by augmenting the initial set with a space-filling design such as the maximum projection (MaxPro) design \citep{joseph2015maximum}. The augmented design can be obtained sequentially by adding one fingerprint at a time using the \texttt{MaxProAugment} function in the R package \texttt{MaxPro} \citep{bamaxpro}. 

We use Gaussian Process (GP) for modeling the energy data. Assume that $e(\cdot)$ is a realization of a GP:
\begin{equation}\label{eq:priorgp}
e(\bm x)\sim GP(\mu,C(\bm x;\cdot)),
\end{equation}
where $\mu$ is the mean and $C(\bm x_u; \bm x_v)=Cov\{f(\bm x_u),f(\bm x_v)\}$ is the covariance function. See \cite{santner2018space}  for details on GP modeling. Given the energy-data, the posterior distribution of $e(\bm x)$ is given by
\begin{equation}\label{eq:posgp}
e(\bm x)|\bm e\sim \mathcal{N}(\hat{e}(\bm x), s^2(\bm x)),
\end{equation}
where 
\begin{equation}\label{eq:estimated_energy}
\hat{e}(\bm x)=\mu+C(\bm x;\bm S)C^{-1}(\bm S;\bm S)(\bm e-\mu\bm 1)     
\end{equation}is the surrogate model, 
\begin{equation}\label{eq:std_energy}
s^2(\bm x) = C(\bm x;\cdot)-C(\bm x;\bm S)C^{-1}(\bm S;\bm S)C(\bm S;\cdot)
\end{equation} is the variance, $\bm S=[\bm x_1^T,\cdots,\bm x_{n}^T]^T$, 
$C(\bm x;\bm S)$ is the covariance vector with $i$th element $C(\bm x;\bm S_i)$, $C(\bm S;\bm S)$ is the covariance matrix, and $\bm 1$ is a vector of 1's. We use the R package \texttt{DiceKriging} \citep{roustant2012dicekriging} to fit the GP model using a Gaussian covariance function. Figure \ref{fig:branin1} (right) shows the potential energy predictions (shown in color) based on the fitted GP model. We see that the estimated global minimum seems to lie in the ``interior'' of the candidate set of configurations. In this case, there is no need to further expand the ``low-energy'' region, as it is already surrounded by the candidate set of configurations. 

\begin{figure}[h]
\begin{center}
\includegraphics[scale=0.28]{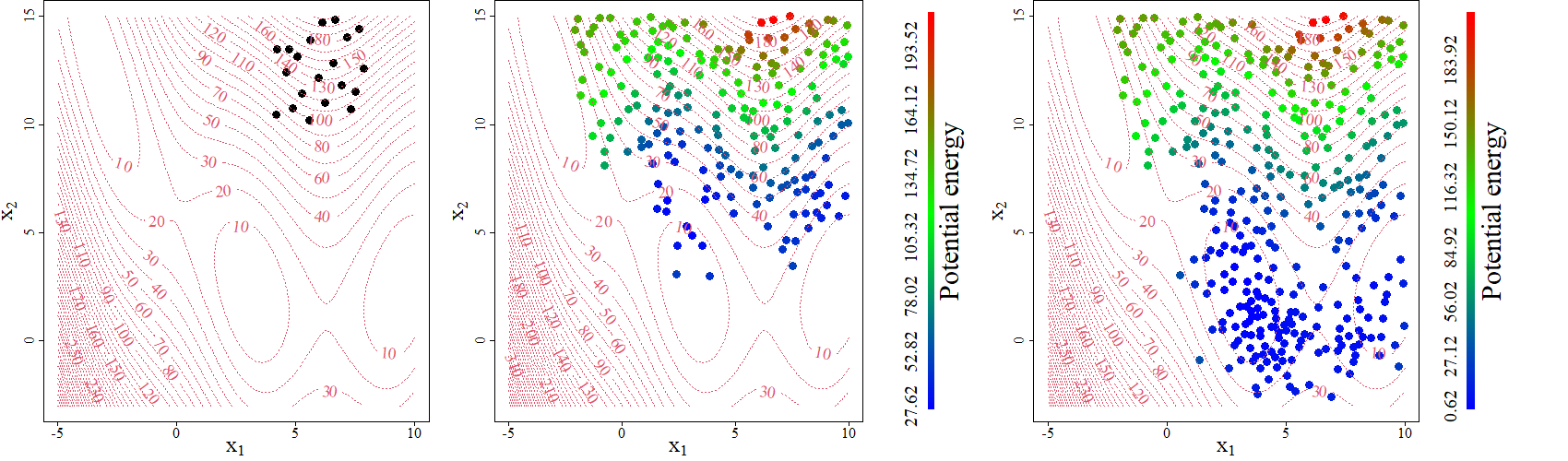}
\caption{Candidate set of  $n_0=20$ initial fingerprints (left); $N_1=200$ fingerprints obtained from the non-adaptive expansion algorithm (center);  and $N_2=341$ fingerprints obtained after the adaptive expansion algorithm (right), shown over the contour plot of the Branin function.}
\label{fig:branin2}
\end{center}
\end{figure}

Now, let us consider another scenario, where the the initial set of $10p$ fingerprints are as shown over the contour plot of the Branin function in 
Figure \ref{fig:branin2} (left). Figure \ref{fig:branin2} (center) shows the non-adaptive expansion along with the potential energy (shown in color) based on the fitted GP model. The estimated global minimum seems to lie at the ``boundary'' of the candidate set of configurations. In this case the ``low-energy'' region is not well explored on all sides. Thus, in this case, we need to further expand the ``low-energy'' region to ensure that the minimum of the ``low-energy'' region, which may potentially be the global minimum or a reliable local minimum, is included in the candidate set of configurations.

The definition of the ``boundary'' and the ``interior'' of the candidate set of configurations is based on the dimension $p$ of the fingerprint. A two-dimensional fingerprint, assumed to be at the origin,  lies in the interior of the domain space if it has neighboring fingerprints in each of the four quadrants, within a distance $r$ around it. Otherwise, it lies on the boundary of the domain space spanned by the candidate set of fingerprints. Here $r$ is taken to be the maximum distance to the nearest neighbor for the candidate set of $N_1$ fingerprints obtained at the end of the non-adaptive expansion algorithm.  Figure \ref{fig:bd_ex} in the Appendix shows examples of two-dimensional fingerprints that lie on the boundary or in the interior of the domain space spanned by the candidate set.


For the case shown in Figure \ref{fig:branin2} (center), we push the boundary of the low-energy region by perturbing the lowest energy fingerprint that lies on the boundary of the spanned domain space. With the addition of every 10 fingerprints (thumb-rule) to the candidate set, DFT is used for computing the potential energy of the fingerprint with the least energy estimate. The GP model is then updated to better estimate the energy in the newly explored lower-energy domain space. A periodic model-update helps navigate the expansion of the lower-energy region. If the fingerprint having the minimum estimated potential energy does not change within $10$ successive DFT computations (thumb-rule), we stop the algorithm. The adaptive domain space exploration algorithm is included as Algorithm 2.

\begin{algorithm}
  \caption{: Adaptive domain space expansion}
  \begin{algorithmic}[1]
  \STATE Import $\mathcal{C}, \mathcal{X}, \bm R, \bm D, t$
  \COMMENT{Obtained at the end of the non-adaptive expansion procedure}
  \STATE $\mathcal{X}_{} \leftarrow \{\bm x_1,\cdots,\bm x_{n_0} \}$
  \STATE Augment $\mathcal{X}_{n_0}$ by $10p-n_0$ space-filling fingerprints to obtain $\mathcal{X}_{DFT}$
  \COMMENT{\texttt{R} package:\texttt{MaxPro} }
  \STATE $\bm e \leftarrow DFT(\mathcal{X}_{DFT})$; model $\leftarrow GP(\mathcal{X}_{DFT},\bm e)$
  \STATE $\hat{\bm e} \leftarrow GP.predict(\mathcal{X})$
  \STATE flag $\leftarrow 1$; DFT\_period $\leftarrow 0$; iter $\leftarrow$ 0;
  \WHILE{flag $= 1$}
  \STATE Find $\bm c_{per}$, the configuration with minimum estimated energy lying on the boundary
  \STATE Lines $8-10$ from the non-adaptive domain space expansion algorithm
  \IF{$d_{min}>t$} 
  \STATE Lines $12-15$ from the non-adaptive domain space expansion algorithm
  \STATE DFT\_period $\leftarrow$ DFT\_period+1
  \STATE $e_{new} \leftarrow GP.predict(\mathcal X_{new})$
  \STATE $\hat{\bm e} \leftarrow$ append($\hat{\bm e},e_{new}$)
  \IF{DFT\_period $=10$}
  \STATE iter $\leftarrow$ iter+1
  \STATE Find $\bm c_{min}$, the configuration with the minimum estimated potential energy
  \STATE $e_{min} \leftarrow DFT(\bm c_{min})$; $\bm e \leftarrow$ append($\bm e,e_{min}$)
  \STATE $\mathcal{X}_{DFT} \leftarrow$
  append($\mathcal{X}_{DFT},\bm c_{min}$)
  \STATE model $\leftarrow GP(\mathcal{X}_{DFT},\bm e)$
  \STATE $\hat{\bm e} \leftarrow GP.predict(\mathcal{X})$
  \IF{iter $= 10$}
  \STATE iter $= 0$
  \IF{$\bm c_{min}$ has not changed since the last 10 DFT computations}
  \STATE flag $= 0$
  \ENDIF
  \ENDIF
  \ENDIF
  \ENDIF
  \STATE $t \leftarrow 0.5(t+d_{min})$
  \ENDWHILE
  \STATE Output $\mathcal C, \mathcal X, \mathcal{X}_{DFT}, \bm e$
  \end{algorithmic}
\end{algorithm}


Figure \ref{fig:branin2} (right) shows the result of applying the adaptive expansion algorithm to the scenario presented in Figure \ref{fig:branin2} (center). The algorithm adaptively expands the set of $N_1=200$ fingerprints obtained at the end of non-adaptive expansion to $N_2=341$ fingerprints. Note that the algorithm continues to expand until the low-energy region is fully explored, and the minimum is well surrounded by the candidate set of fingerprints.


We tested our algorithms (non-adaptive + adaptive expansion) to find the minimum of two different kinds of functions - sphere function and Schwefel function \citep{simulationlib}. The sphere function is simple and smooth with only one minimum, while the Schewefel function is complex with several local minima. These functions are generalizable to any dimension $p$. We consider two distinct values of dimension: $p=2$ and $p = 10$ for both the functions. We also consider two distinct values of budget (until the non-adaptive expansion step): $N_1 = 50p$ and $N_1=100p$. 

The feasible domain space of the sphere function is taken as $[-5.12,5.12]^p$ and that of the Schwefel function to be $[-500,500]^p$.
However, as the feasible domain space is assumed to be unknown, we generate the initial set of $10p$ fingerprints as a maximin Latin hypercube design \citep{morris1995exploratory} in a smaller sub-space: $[1.5, 4]^p$ for the sphere function and  $[250, 400]^p$ for the Schwefel function. Note that the global minimum of the sphere function, $(0,...,0)$, and that of the Schewefel function, $(420.9687,...,420.9687)$, are outside these sub-spaces. We compare our results to a baseline maximin Latin hypercube sample of size $N_2$ generated using  the R package \texttt{lhs} \citep{carnell2016package} in the sub-space of the initial set of fingerprints.
We perform 30 simulations for each unique combination of function, dimension ($p$), and budget until non-adaptive expansion ($N_1$). 

\begin{figure}[h]
\begin{center}
\includegraphics[scale=0.42]{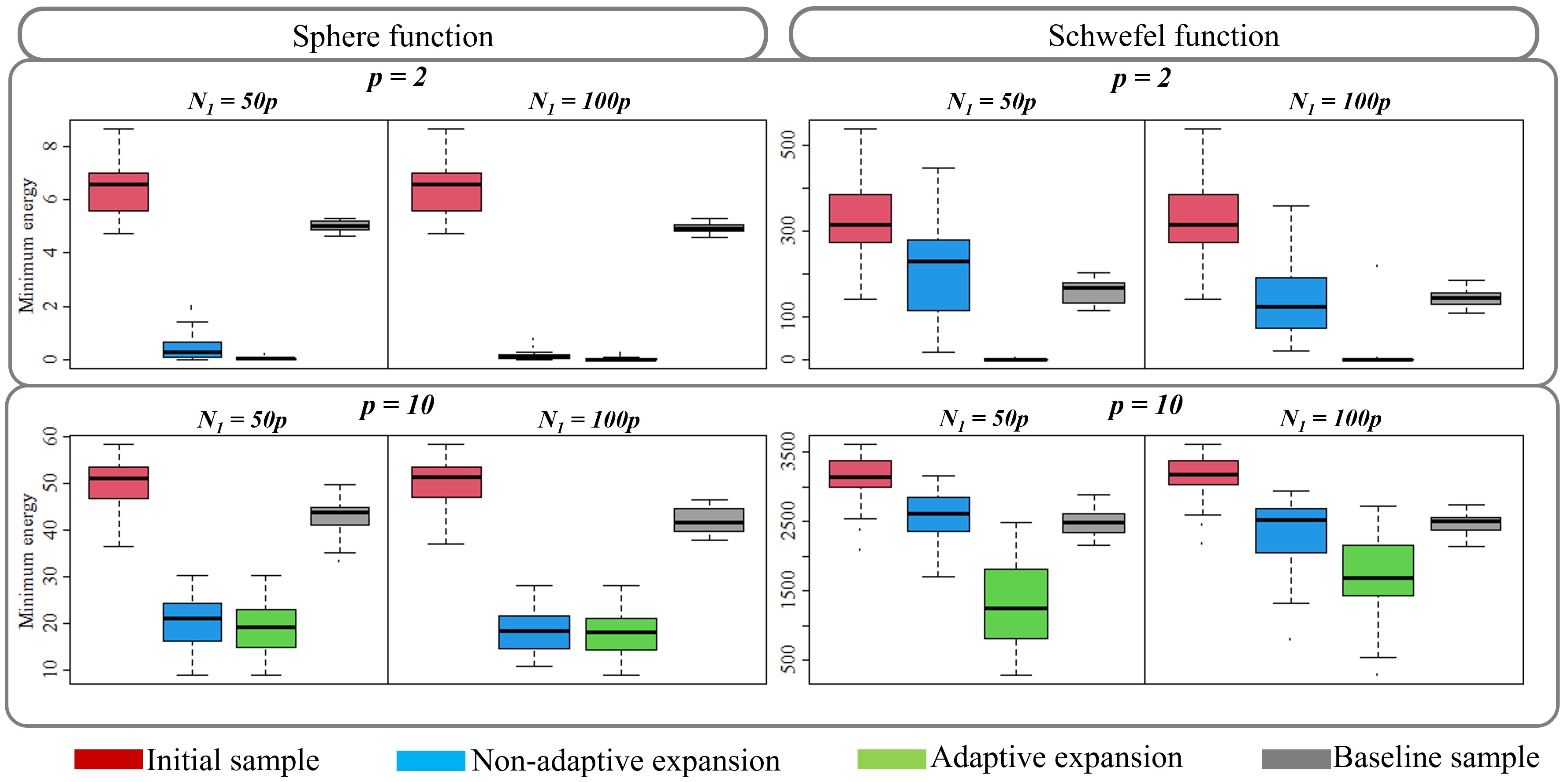}
\caption{Distribution of the minimum energy in the candidate set of fingerprints obtained at each step of our method (initial sample, non-adaptive expansion, and adaptive expansion) and in the baseline sample.}
\label{fig:simulations}
\end{center}
\end{figure}

Figure \ref{fig:simulations} shows the distribution the minimum energy in the candidate set of fingerprints obtained at each step of our method, and in the baseline sample. There are several important points to note. First, the candidate set generated by our method (at the end of the adaptive expansion step) has a fingerprint with a lower energy, on average, as compared to the maximin Latin hypercube design (baseline sample) of the same size. Second, the adaptive expansion step of our algorithm helps navigate the search towards lower energy regions of the feasible space. Third, a larger budget in the non-adaptive expansion step leads to more exploration of the feasible domain space, which results in lower energy fingerprints in the candidate set, as expected. Fourth, our method scales up well, performing better than the state-of-the-art method even for higher dimensions. Fifth, our method works well even for very complex and highly multimodal functions.

\subsection{Exploration and exploitation of the domain space: Bayesian optimization}
The purpose of this step is to identify the crystal structure configuration with the least potential energy in this candidate set of size $N_2$ obtained from the expansion steps. As mentioned in Section $3.3$, energy computation using the DFT is too expensive, which makes it impractical to compute the energy for all the $N_2$ configurations. So, we will use the GP model developed during the adaptive expansion procedure to estimate the energy of all the $N_2$ fingerprints. Bayesian optimization \citep{jones1998efficient, frazier2018} will then be used for iteratively optimizing and updating the model. The method will let us identify the global minimum with DFT computations over only a small fraction of the $N_2$ fingerprints in the candidate set.

Let $\bm e$ be the vector of potential energy, computed using DFT, and $s(\cdot)$ be the standard error of their energy estimate.  Then, the expected improvement criterion can be expressed as the following closed form \citep{jones1998efficient}:
\begin{equation}\label{eq:ei2}
EI(\bm x) = [\min(\bm e)-\hat{e}(\bm x)]\Phi\bigg(\frac{\min(\bm e)-\hat{e}(\bm x)}{s(\bm x)}\bigg)+s(\bm x)\phi\bigg(\frac{\min(\bm e)-\hat{e}(\bm x)}{s(\bm x)}\bigg),
\end{equation}
where $\Phi$ and $\phi$ are respectively the cumulative distribution function and the probability density function of the standard normal distribution, $\hat{e}(.)$ is the estimated energy as defined in \eqref{eq:estimated_energy}, and $s(\bm x)$ is as defined in \eqref{eq:std_energy}. As the surrogate model is cheap, we evaluate \eqref{eq:ei2} on all the $N_2$ fingerprints in the candidate set $\mathcal{X}$, and find the one that maximizes it:
\begin{equation}\label{eq:fnew}
\bm x_{new} = \argmax_{\bm x \in \mathcal{X}} EI(\bm x).
\end{equation}
DFT is used for computing the potential energy at $\bm x_{new}$, and the energy-data set is updated to include [$\bm x_{new},e(\bm x_{new})$]. Then, we use \eqref{eq:posgp} to update our surrogate model based on the updated energy-data. The expected improvement (EI) criterion balances exploration of the PES with exploitation, thereby simultaneously addressing both the objectives of crystal structure prediction - exploiting low-energy regions to search for the minimum and exploring new and unusual domains of the PES. We stop the Bayesian optimization algorithm, when the expected improvement becomes negligible as compared to the current estimate of the minimum potential energy. The Bayesian optimization algorithm is included as Algorithm 3.

\begin{algorithm}
  \caption{: Bayesian optimization}
  \begin{algorithmic}[1]
  \STATE Import $\mathcal{C},\mathcal{X}$
  \COMMENT{Obtained from the expansion algorithms}
  \STATE Input $t_{EI}$
  \IF{Algorithm 2 : Adaptive domain space expansion is used}
  \STATE Import $\mathcal{X}_{DFT}, \bm e$
  \ELSE
  \STATE Lines $2-5$ from the adaptive domain space expansion algorithm
  \ENDIF
  \STATE $e_{min} \leftarrow \min(\bm e)$
  \STATE $EI_{i} \leftarrow EI(\bm x_i); i \in \{1,\cdots,N_2\}$ \COMMENT{Use \eqref{eq:ei2}}
  \STATE percent\_improve $\leftarrow \max(\bm{EI})/e_{min}$
  \WHILE{percent\_improve $\le t_{EI}$}
  \STATE $i\_maxEI \leftarrow \argmax_i (\bm{EI}); i \in \{1,\cdots,N_2\}$
  \STATE $e_{new} \leftarrow DFT(\bm x_{i\_maxEI})$
  \STATE $\mathcal{X}_{DFT} \leftarrow$ append($\mathcal X_{DFT}, \bm x_{i\_maxEI}$)
  \STATE $\bm e \leftarrow$ append$(\bm e,e_{new})$
  \STATE $model \leftarrow GP(\mathcal X_{DFT},\bm e)$
  \STATE $EI_{i} \leftarrow EI(\bm x_i); i \in \{1,\cdots,N_2\}$ \COMMENT{Use \eqref{eq:ei2}}
  \STATE percent\_improve $\leftarrow \max(\bm{EI})/e_{min}$
  \ENDWHILE
  \STATE $i\_stable \leftarrow \argmin_i \bm e; i \in \{1,\cdots,nrows(\mathcal X_{DFT})\}$
  \STATE Output $\bm x_{i\_stable}, \bm c_{i\_stable}$
  \end{algorithmic}
\end{algorithm}

The computational complexities of the three Algorithms 1, 2, and 3 are 
$O((n_0+n_1)^2p^2)$, $O(n_0+n_2/10)^3n_2)$, and $O((n_0+n_2/10+n_3)^3n_3)$, respectively.
See Section B of the Appendix for details. In addition to computational costs, there are storage costs as it is required to store $N_2$, $p$-dimensional fingerprints, their corresponding Cartesian coordinate configurations, and the vectors consisting of the estimated energy and distance to the nearest neighbor of each fingerprint. However, the main cost of the methodology is the cost of running $(n_0+n_2/10+n_3)$ DFT computations.

\section{Example: Crystal structure of $Al_8$}

The objective is to find the most stable crystal structure configuration of $Al_8$, or the configuration of eight Aluminum (Al) atoms arranged in a parallelepiped unit cell. Note that the most stable configuration of $Al_8$ is the face-centered cubic (fcc), which is already known  and is shown in Figure \ref{fig:AL_true_structure}. Because the primitive cell of fcc $Al$ has one atom, it must be included in the configuration space of $Al_8$. For our candidate set, eight-atoms structures were created by (1) randomly choosing a specific value of volume $v$ falling within $\pm 5\%$ of the known specific volume of the Al fcc structure, (2) randomly selecting three vectors $\vec a$, $\vec b$, and $\vec c$ of the unit cell so that its volume, given by $V \equiv {\vec a}\cdot(\vec b\times \vec c) = 8v$, and (3) randomly arranging the eight Al atoms in the cell so that the distance between any pairs is larger than $2.0$ \AA. Two constraints (1) and (3) of this procedure, which were formulated from the known facts of the fcc Al structure, clearly limit the examined configuration space but the search domain remains staggering and certainly contains the global minimum.

Within our expansion-exploration-exploitation framework, we start from a set of $270$ initial configurations for which the energy is computed at the DFT level. Note that the DFT calculations performed herein involve {\it only single-point energy calculations} but not any local optimizations, which may be $10^3-10^4$ times more expensive. Therefore, in general, none of the examined structures is a local minimum of the PES. However, the main objectives of this work, i.e., {\it (diversely) filling the configuration space and searching for the global minimum of a big and diverse structure dataset}, can be demonstrated and is very useful for material structure prediction.

To obtain a candidate set of fingerprints, we start with the non-adaptive domain space  expansion algorithm (Algorithm 1). We have a set of $n_0=270$ initially known atomic configurations of $Al_8$. These configurations become the input $\mathcal C$, and their corresponding fingerprints become the input $\mathcal X$. The fingerprints have a dimension of $p=32$. As per the thumb-rule mentioned earlier, we take $N_1= 100p = 3,200$. For perturbing the most sparsely located atomic configuration, we randomly change the position of each atom by a maximum of 0.1 Angstrom. 

\begin{figure}[h]
\begin{center}
\includegraphics[scale=0.35]{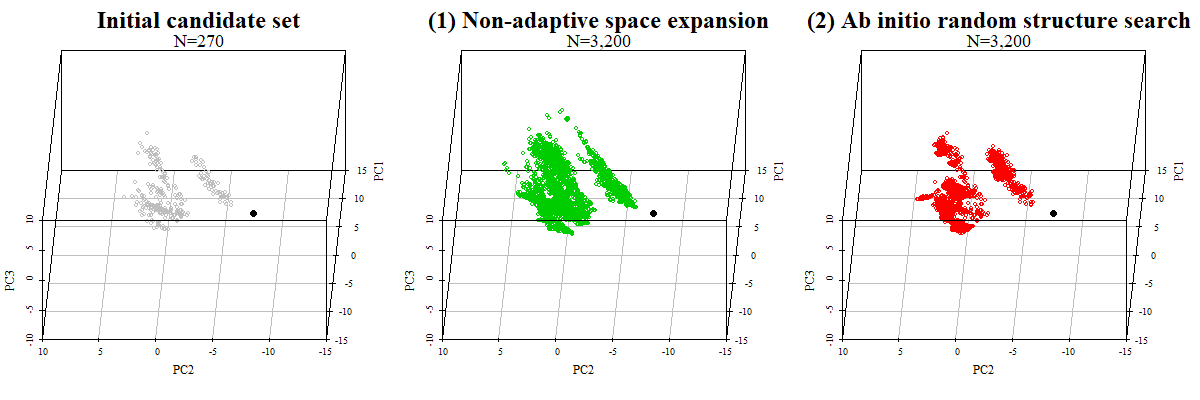}
\caption{Initial set of fingerprints (left), and  candidate set of $N_1=3,200$ fingerprints obtained using non-adaptive space expansion algorithm (center) and the Ab initio random structure searching approach (right). The solid black circle is the true global optimum, which is not included in the candidate set.}
\label{fig:real_ex_fp}
\end{center}
\end{figure}

As the $AGNI$ fingerprint is $32$-dimensional, the candidate set of fingerprints cannot be visualized directly. We use Principal Component Analysis \citep{jackson2005user}, or PCA to visualize the first three PCs, which capture $97\%$ of their variance. Figure \ref{fig:real_ex_fp} (left) shows the PCs of the initial set of fingerprint (grey circles) that are input to the non-adaptive space expansion algorithm (Algorithm 1). The solid black circle (in this and all the subsequent figures) corresponds to the most stable fingerprint, or the fingerprint that has the minimum potential energy. This is not a part of the initial candidate set. However, this is the solution that we hope to achieve. Note the relatively large gap between it and the initial set of fingerprints. Ideally, our expansion algorithms will expand the initial candidate set to include the most stable fingerprint, and then the Bayesian optimization algorithm should identify this fingerprint as the global minimum.

Figure \ref{fig:real_ex_fp} (center) shows the PCs for the candidate set of $N_1=3,200$ fingerprints obtained using our non-adaptive space expansion algorithm. Note that the algorithm expands the volume of the domain space spanned by the initial candidate set of fingerprints, which reduces the gap between different clusters of fingerprints in the initial candidate set.

\begin{figure}[h]
\begin{center}
\includegraphics[scale=0.6]{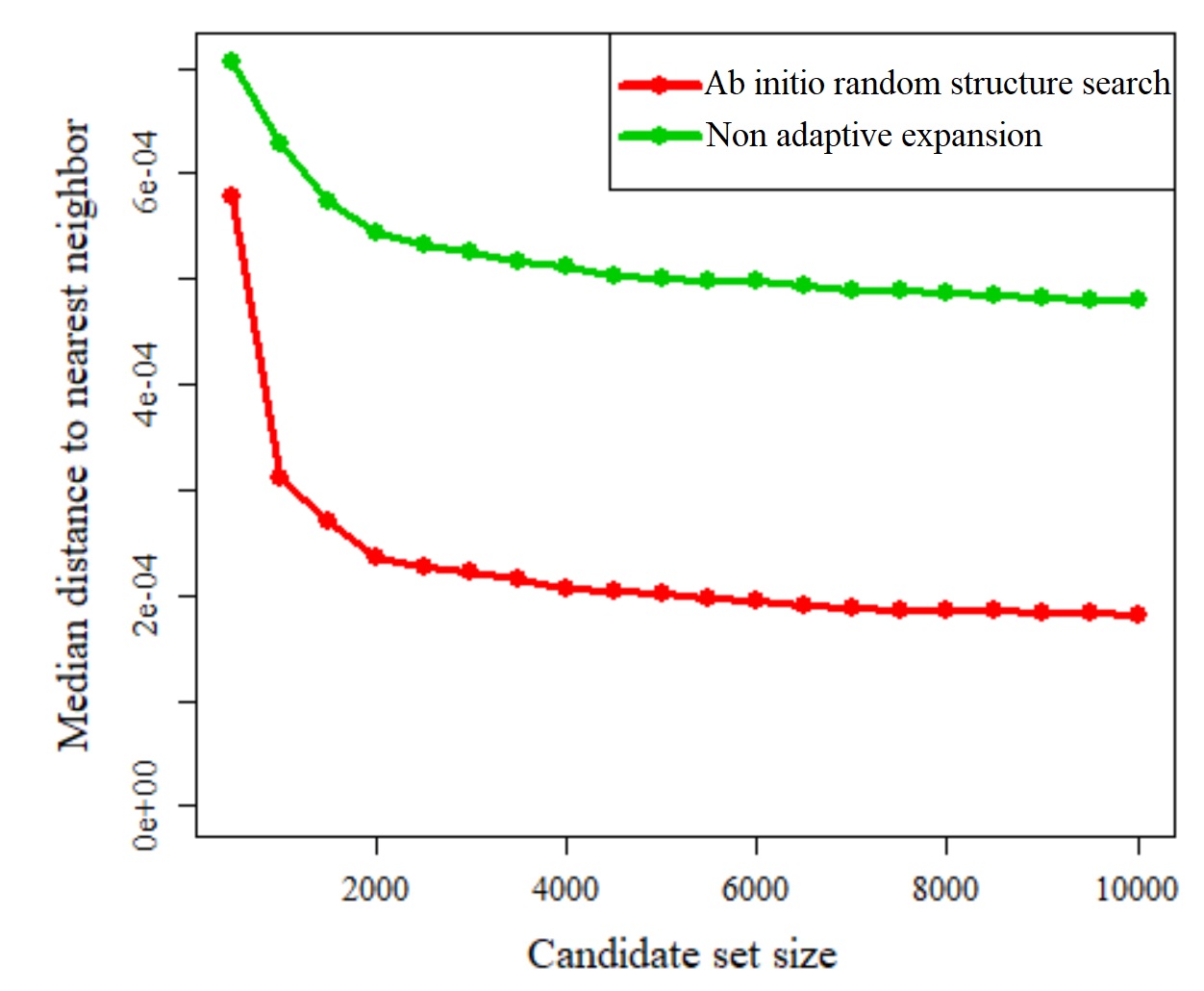}
\caption{Distance to the nearest neighbor vs candidate set size for fingerprints.}
\label{fig:nearest_nb_dist}
\end{center}
\end{figure}

Figure \ref{fig:real_ex_fp} (right) shows the PCs for the candidate set of $N_1=3,200$ fingerprints obtained using the Ab initio random structure searching approach (AIRSS) \citep{pickard2011ab}. 
Figure \ref{fig:nearest_nb_dist} shows the median distance to the nearest neighbor of a fingerprint as the candidate set size increases. As expected, our proposed candidate set keeps expanding the spanned domain space leading to more spaced-out fingerprints. On the other hand, the Ab initio random structure searching approach results in a set of closely packed fingerprints leading to wastage of resources, and failure to consider more distinct configurations.

\begin{figure}[h]
\begin{center}
\includegraphics[scale=0.4]{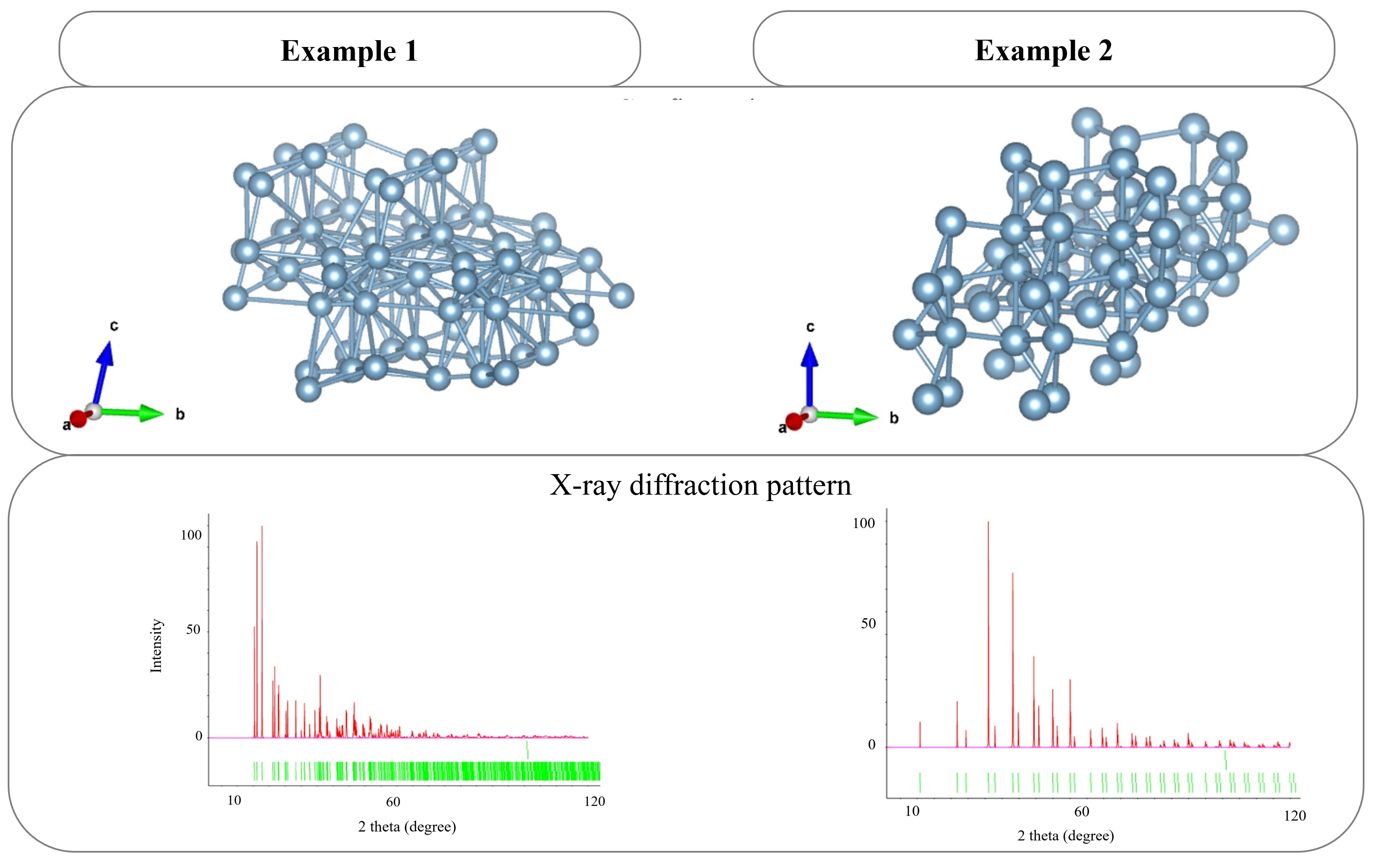}
\caption{Configuration (top) and X-ray diffraction pattern (bottom) of a couple of example structures.}
\label{fig:Example_configs}
\end{center}
\end{figure}

Figure \ref{fig:Example_configs} (top) shows examples of a couple of configurations corresponding to the fingerprints obtained from the non-adaptive domain space expansion algorithm. Figure \ref{fig:Example_configs} (bottom) shows the X-ray diffraction patterns corresponding to these  configurations. The stark differences in these configurations, as evident by their X-ray diffraction pattern, shows that the algorithm spans through quite distinct regions of the domain space.

After obtaining a candidate set of $3,200$ fingerprints from the non-adaptive expansion algorithm, we compute the energy for $10p=320$ fingerprints, using DFT, to develop a GP model. These $320$ fingerprints include the initial candidate set of $270$ fingerprints corresponding to known configurations, which are augmented by another $50$ fingerprints from the candidate with the space-filling MaxPro design. Figure \ref{fig:maxpro_real_ex} in the Appendix shows the initial fingerprints and the ones augmented using the MaxPro design. 

\begin{figure}[h]
\begin{center}
\includegraphics[scale=0.4]{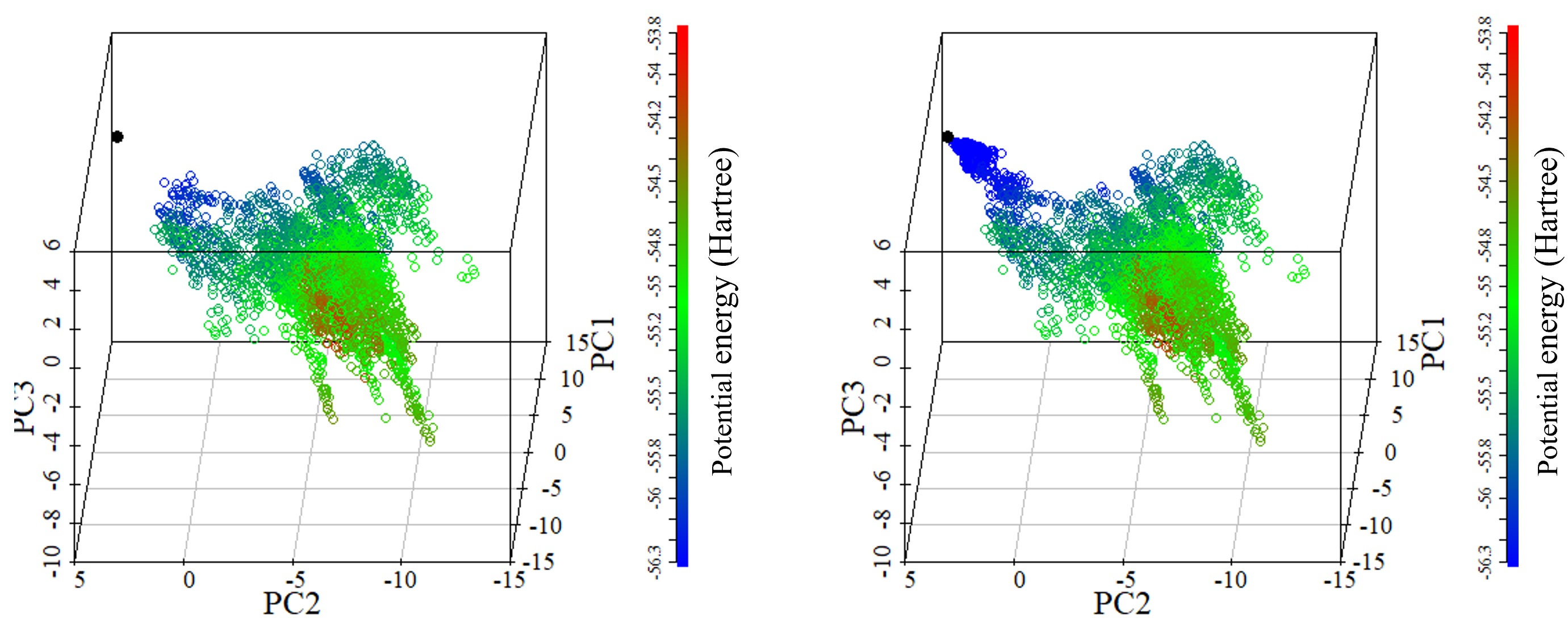}
\caption{Potential energy of the candidate set of fingerprints obtained after non-adaptive expansion (left) and adaptive expansion (right).}
\label{fig:pe_contour_real_ex}
\end{center}
\end{figure}

Figure \ref{fig:pe_contour_real_ex} (left) shows the predicted potential energy  of the candidate set using the GP model. The low-energy regions seem to be around the boundary of the spanned domain space. Using the boundary definition mentioned earlier and illustrated in Figure \ref{fig:bd_ex}, we find that the estimated minimum of the candidate set is actually at the boundary of the spanned domain space. So, it is necessary to further expand this low energy region as it may lead to further minimization of the current estimate of the energy-minimum. Note that we work in the space of the first three principal components for determining the boundary because it is very expensive to work in the $32$-dimensional fingerprint space.

The adaptive expansion algorithm (Algorithm 2) is used for further expanding the identified low-energy region of the spanned domain space. Figure \ref{fig:pe_contour_real_ex} (right) shows the updated candidate set of fingerprints and the  predicted potential energy after the adaptive expansion procedure. The algorithm adds $530$ fingerprints to the candidate set, and DFT computations are done for every $10th$ fingerprint added to the set, i.e., for a total of $53$ fingerprints. The algorithm stops when the estimated minimum does not change with $10$ successive DFT computations. We observe that the algorithm succeeds in expanding the candidate set towards the unknown true global minimum!

Once a candidate set of fingerprints is obtained by the expansion algorithms, we explore and exploit it for the global minimum of potential energy, using the Bayesian optimization algorithm (Algorithm 3). The input $\mathcal X$ in this example is the candidate set of $3,200+530 = 3,730$ fingerprints that we obtained using the expansion algorithms.

\begin{figure}[h]
\begin{center}
\includegraphics[scale=0.4]{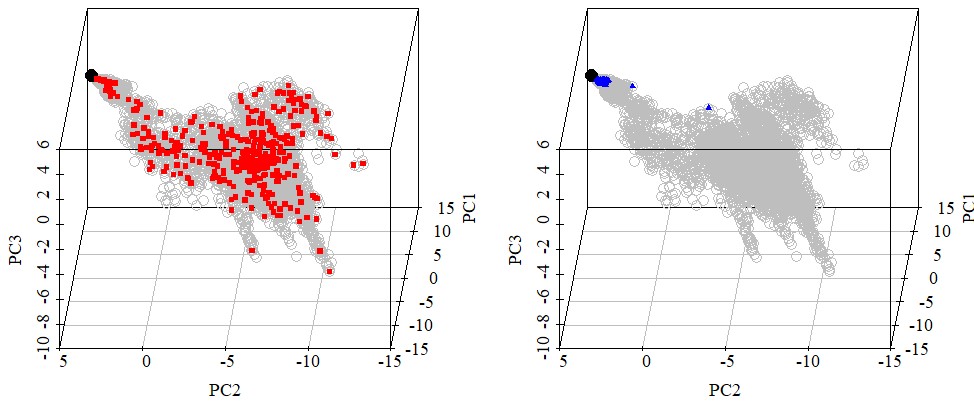}
\caption{(left): initial set of $373$ fingerprints for which the potential energy is computed using DFT (red squares) and (right): $95$ fingerprints selected during the Bayesian optimization (blue triangles), which are shown over the candidate set of $3,200$ fingerprints (circles).} 
\label{fig:BO_optim}
\end{center}
\end{figure}

The circles in Figure \ref{fig:BO_optim} are the candidate set of $N_1=3,200$ fingerprints. The potential energy was computed for a set of $320+53=373$ fingerprints (in the adaptive domain space expansion algorithm), shown as squares in Figure \ref{fig:BO_optim} (left). A GP model was fitted using the potential energy data of the $373$ fingerprints. Then, iterative Bayesian optimization procedure of updating the GP model, and adding a point in the known-data based on the EI criterion is continued until the expected improvement becomes lesser than a threshold value $t_{EI}$. We chose $t_{EI}$ to be $0.001\%$ of the current minimum value of potential energy in the energy-data. Thus, for the $ith$ iteration, the threshold value is:
\begin{equation}\label{eq:ei_threshold}
t_{EI} = 10^{-5} \times \min(e_1,\cdots,e_{n+i}).
\end{equation}
This is a reasonably low value as the potential energy in the known-data varies by $5\%$ around its mean. \cite{jones1998efficient} stop the Bayesian optimization algorithm when the expected improvement is 1\% of the current best function value. We have used a much more conservative stopping criterion. For a different structure, a reasonably low cutoff must be chosen depending on the variance in the potential energy of the candidate set of configurations. With the above threshold, the algorithm stopped after the $95th$ iteration. 

The fingerprints iteratively added in the known-data are shown as blue triangles in Figure \ref{fig:BO_optim} (right). There are three points to note about the iteratively added fingerprints. First, $93$ of the $95$ fingerprints are selected in the region around the global minimum, which shows that the algorithm does well in exploiting the ``promising'' region of the domain space. Second, two fingerprints are selected in regions far away from the global minimum. These are regions at the boundary of the domain space, where probably there is high uncertainty in the potential energy estimate of the GP surrogate model. This shows the exploratory nature of the algorithm, where it tries to find the global minimum in regions other than the ``promising region''. This exploratory feature of the algorithm makes it better than the state-of-the-art approaches such as basin hopping \citep{wales1997global}  and minima hopping \citep{goedecker2004minima}, which focus only on exploiting the ``promising region'' of the domain space for the global minimum. Third, though the true global minimum was not a part of the candidate set, we identified the fingerprint closest to it as the global minimum! Thus, the algorithm provided a solution that is potentially very similar to the true global minimum. The expected improvement criterion, and the energy of configurations selected for DFT computations are visualized in Figure \ref{fig:EI_analysis} in the Appendix.

\begin{figure}[h]
\begin{center}
\includegraphics[scale=0.45]{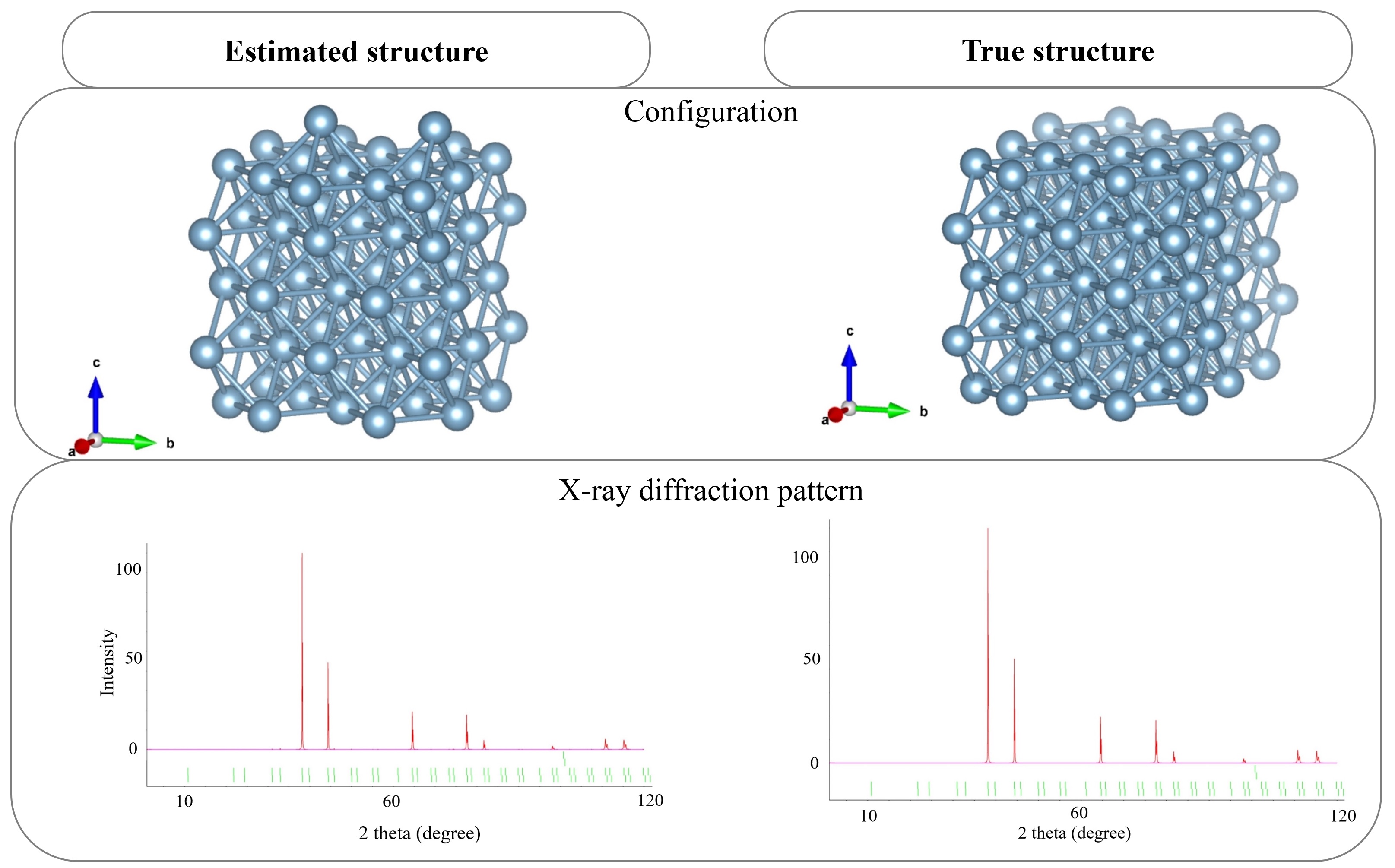}
\caption{Configuration (top) and X-ray diffraction pattern (bottom) of the estimated and true structures.} 
\label{fig:result_comparison}
\end{center}
\end{figure}

The algorithm's output and our solution is the configuration corresponding to the fingerprint selected in the $368th$ DFT computation, as it has the minimum potential energy in the energy-data. Figure \ref{fig:result_comparison} (top) shows that the estimated structure looks quite similar to the true structure. The similarity in the estimated and true structures is more evident in the X-ray diffraction pattern of the structures as shown in Figure \ref{fig:result_comparison} (bottom). 


\section{Conclusion}

We have developed an active learning method to (1) obtain a candidate set of crystal structure configurations that expands the space of a few initially known configurations, and (2) find the configuration with the lowest potential energy in the set. The novelty of our approach is the expansion-exploration-exploitation framework that extends the traditionally used exploration-exploitation Bayesian optimization framework to better achieve the objectives of crystal structure prediction. The expansion algorithms ensure that the candidate set continues to expand with the addition of each fingerprint - first in arbitrary directions (with respect to potential energy) to explore new and possible unusual domains of the PES, and then towards lower-energy atomic configurations. Our algorithm provides a space-filling design without the knowledge of the boundaries of the design space. This is a novel contribution to the field of experimental design, where most of the work on space-filling design is focused on cases of known design space.

Although we demonstrated our approach on a simple problem of finding the stable configuration of $Al_8$, the new concepts are powerful and can easily be generalized to more realistic problems. A recent and interesting application of the proposed method in inverse designs can be found in \cite{krishna2022inverse}. 

\medskip

\section*{\Large{Acknowledgments}}
The authors thank U.S. National Science Foundation grants DMREF-1921873 and XSEDE through grant DMR-170031.

\medskip

\bibliography{Ref_all}

\section*{\LARGE{Appendices}}

\setcounter{figure}{0}
\appendix


\section{Additional Figures}

\begin{figure}[h]
\begin{center}
\includegraphics[scale=0.75]{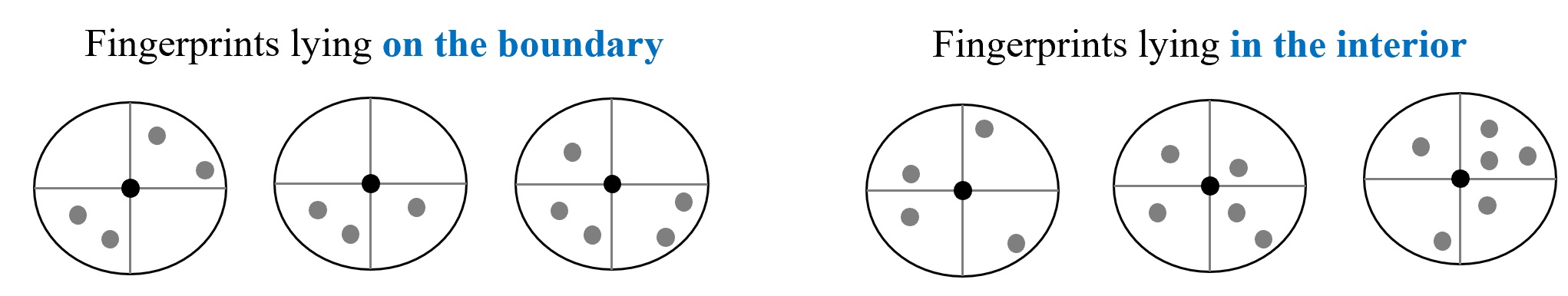}
\caption{Examples of two-dimensional fingerprints that lie (left): on the boundary of the domain space spanned by the candidate set; (right): in the interior of the domain space spanned by the candidate set. Note that the radius of the circle is $r$.}
\label{fig:bd_ex}
\end{center}
\end{figure}

Figure \ref{fig:bd_ex} illustrates the definition of ``boundary'' and ``interior'' in a two-dimensional example.

\begin{figure}[h]
\begin{center}
\includegraphics[scale=0.5]{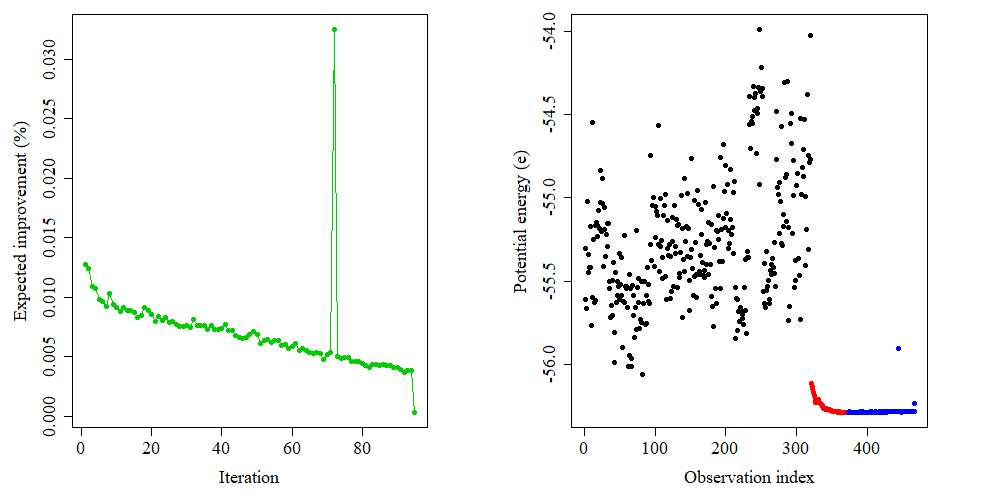}
\caption{(left): Percentage of expected improvement with respect to the current energy minimum estimate; (right): Potential energy of all the fingerprints in the known-data - for the initial known fingerprints (in black), for the fingerprints added by adaptive expansion (in red), for the fingerprints added during the Bayesian optimization procedure (in blue).} 
\label{fig:EI_analysis}
\end{center}
\end{figure}

Figure \ref{fig:EI_analysis} (left) shows the expected improvement as a percentage of the current minimum estimate of potential energy from the energy-data. The expected improvement has a decreasing trend with the number of iterations. As the algorithm learns the potential energy surface and exploits promising locations for global minimum, a lesser improvement in the current global minimum estimate is expected in further iterations. Figure \ref{fig:EI_analysis} (right) shows the potential energy of the first $n=320$ observations of the known-data (black circles), followed by that of the $53$ fingerprints (red circles) iteratively added to the known-data during the adaptive expansion procedure, which are in-turn followed by the $95$ fingerprints (blue) added to the known data during the Bayesian optimization procedure. This figure makes it clear that the non-adaptive expansion algorithm expands the domain space without considering the potential energy, the adaptive expansion algorithm drives the expansion towards lower-energy regions, and the Bayesian optimization procedure explores the candidate set and exploits the low-energy regions for the global minimum of potential energy.

\begin{figure}[h]
\begin{center}
\includegraphics[scale=0.45]{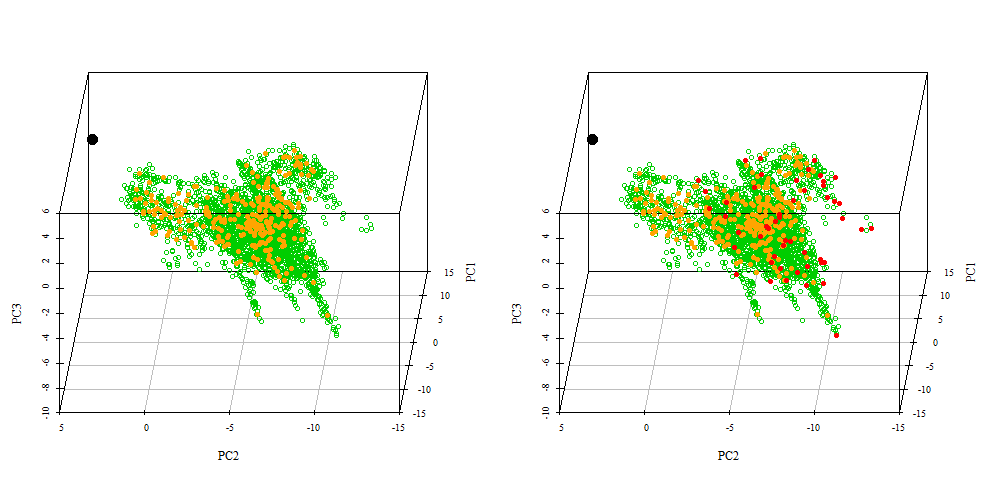}
\caption{(left): Initial set of fingerprints (orange) over the candidate set of fingerprints (green); (right): Initial set of fingerprints augmented by the space-filling MaxPro design (red).}
\label{fig:maxpro_real_ex}
\end{center}
\end{figure}

\newpage


\section{Computational complexity}
Here are the details of the computational complexity of our methodology. The non-adaptive expansion step has a complexity of $O((n_0+n_1)^2p^2)$ as it requires computation of $N_1 = n_0+n_1$, $p$-dimensional fingerprints' Euclidean distance to their nearest neighbor in each of their $2p$ neighborhoods.

The computational complexity of fitting a GP over $k$ points is $O(k^3)$. In the adaptive expansion step, the first GP is based on $n_0$ fingerprints (assuming $n_0=10p)$, and is updated after the addition of every 10 fingerprints to the candidate set. Thus the last GP is based on $(n_0+n_2/10)$ fingerprints. This leads to a complexity of $O((n_0+n_2/10)^3n_2)$ for the adaptive expansion procedure. As in non-adaptive expansion, the adaptive expansion algorithm has a complexity of $O(((n_0+n_1)n_2+n_2^2/2)p^2)$ associated with Euclidean distance computation to the nearest neighbor for each fingerprint added to the candidate set. However, since this cost is negligible compared to the cost of fitting the GP model, it is ignored.

In case of Bayesian optimization, a GP is fit repeatedly until the expected improvement becomes negligible. The first GP is based on $(n_0+n_2/10)$ fingerprints, while the last one is based on $(n_0+n_2/10+n_3)$ fingerprints. This leads to a computational complexity of $O((n_0+n_2/10+n_3)^3n_3)$ for the Bayesian optimization procedure.

\end{document}